\documentclass[twoside,11pt]{article}

\usepackage{obs_study_style}
\usepackage{graphicx}
\usepackage{url}
\usepackage{amsmath}
\usepackage{amssymb}
\usepackage{amsfonts}
\usepackage{listings}
\usepackage{xcolor}

\definecolor{codegreen}{rgb}{0,0.6,0}
\definecolor{codegray}{rgb}{0.5,0.5,0.5}
\definecolor{codepurple}{rgb}{0.58,0,0.82}
\definecolor{backcolour}{rgb}{0.95,0.95,0.92}

\ShortHeadings{Bayesian Semiparametric Causal Inference for Event-Time Outcomes}{Ji and Oganisian}
\firstpageno{1}

\begin{document}

\title{\texttt{causalBETA}: An R Package for Bayesian Semiparametric Causal Inference with Event-Time Outcomes}

\author{\name Han Ji \email han\_ji@brown.edu \\
       \addr Department of Biostatistics\\
       Brown University\\
       Providence, Rhode Island, USA 
       \AND
       \name Arman Oganisian \email arman\_oganisian@brown.edu \\
       \addr Department of Biostatistics\\
       Brown University\\
       Providence, Rhode Island, USA }

\maketitle

\begin{abstract}
     Observational studies are often conducted to estimate causal effects of treatments or exposures on event-time outcomes. Since treatments are not randomized in observational studies, techniques from causal inference are required to adjust for confounding. Bayesian approaches to causal estimates are desirable because they provide 1) prior smoothing provides useful regularization of causal effect estimates, 2) flexible models that are robust to misspecification, 3) full inference (i.e. both point and uncertainty estimates) for causal estimands. However, Bayesian causal inference is difficult to implement manually and there is a lack of user-friendly software, presenting a significant barrier to wide-spread use. We address this gap by developing \texttt{causalBETA} (\textbf{B}ayesian \textbf{E}vent \textbf{T}ime \textbf{A}nalysis) - an open-source \texttt{R} package for estimating causal effects on event-time outcomes using Bayesian semiparametric models. The package provides a familiar front-end to users, with syntax identical to existing survival analysis \texttt{R} packages such as \texttt{survival}. At the same time, it back-ends to \texttt{Stan} - a popular platform for Bayesian modeling and high performance statistical computing - for efficient posterior computation. To improve user experience, the package is built using customized \texttt{S3} class objects and methods to facilitate visualizations and summaries of results using familiar generic functions like \texttt{plot()} and \texttt{summary()}. In this paper, we provide the methodological details of the package, a demonstration using publicly-available data, and computational guidance.
\end{abstract}


\newpage

\section{Introduction}
Across a wide range of application areas, observational studies are often conducted to estimate causal effects of treatments on time-to-event outcomes. For instance, biomedical studies in oncology are often interested in analyzing effects of different chemotherapy agents on time-to-death - leading to a large literature under the ``survival analysis'' umbrella. Labor economists are often interested in analyzing the effect of various policies or job training programs on unemployment spells, leading to a literature on ``duration analysis'' methods for modeling time to re-employment. In these settings, the target estimands of interest - whether explicitly acknowledged or not - are often causal in nature. We typically wish to know, say, the difference in the proportion of the target population surviving more than 3 years had everyone in the population received one treatment over another.

Such analyses are complicated from both a causal identification perspective and a statistical estimation perspective. The former complication arises since the treatment of interest is not randomized in observational studies. This motivates the use of causal methods to tease out the treatment's effect on survival from effects of confounding factors (those which impact both selection into treatment and survival). The statistical complexities arise due to right-censoring of event-times: subjects may be censored either due to closure of the observational data base or dropout before the event of interest is observed. In these settings, we observe only a lower bound on the event time of interest. Moreover, in small samples, estimation becomes difficult if there is heavy censoring and relatively few events observed over follow-up. Inadequately regularized nonparametric estimation typically leads to erratic estimates with poor small-sample operating characteristics.

Popular \texttt{R} packages for right-censored event time analysis such as \texttt{survival} \citep{survival-package}, \texttt{eha}, and \texttt{flexsurv} \citep{flexsurv} provide user-friendly functions for common estimators of survival functions and associated quantities such as Kaplan-Meier estimation, Cox's partial likelihood, as well as maximum likelihood estimators of parametric proportional hazard (PH) and accelerated failure time (AFT) models. While these packages provide simple user-interfaces with familiar syntax, they do not perform causal adjustment. In general, it is well known that regression parameters from hazard and AFT models do not have causal interpretations \citep{hernan2010}.

Frequentist methodology for causal survival analysis have been developed \citep{laan2003,tsiatis2019,murray2021}. These usually involve either inverse-weighting marginal discrete-time failure models or fitting conditional discrete-time failure models followed by a post-hoc g-omputation procedure \citep{Robins1986} to back out the marginal survival probabilities. Doubly robust estimation procedures have also been proposed with asymptotic guarantees \citep{westling2023}. Recently, Bayesian approaches to causal estimation have gained popularity \citep{li2023, Oganisian2021}. This is chiefly due to the existence of Bayesian nonparametric and semiparametric survival models that provide flexibility and robustness to misspecification \citep{ibrahim2001, Roy2016,Hu2022}. Unlike frequentist nonparametric methods, inference is finite-sample and based on the posterior. This simplifies uncertainty estimation with flexible models by removing the need for asymptotic approximations or bootstraps. Moreover, posterior point and interval estimation for functionals of the survival distribution is automatic. That is, a posterior over the nonparametric model governing the survival time distribution induces a posterior over causal contrasts of this distribution. Additionally, priors can be used to stabilize estimates in small sample settings. This is useful because event time analysis in practical settings may involve data with high censoring rates and a low number of events.

While Bayesian estimation of causal effects on survival outcomes may be desirable, the required Markov Chain Monte Carlo (MCMC) computation and posterior g-computation can be difficult to code for analysts without implementation-level knowledge of Bayesian methods. \texttt{Stan} is a general-purpose platform for doing Bayesian inference: it take as input a user-provided likelihood, prior, and data and output a specified number of draws from the posterior distribution of the likelihood parameters. On the back-end, \texttt{Stan} runs a variant of Hamiltonian Monte Carlo (HMC) \citep{neal2011mcmc} to obtain posterior draws. HMC uses Hamiltonian dynamics to explore posteriors more efficiently even when they are high-dimensional, as they typically are with semiparametric models. However, it still requires learning essentially a completely new programming syntax to code the required \texttt{.stan} files. Corresponding wrappers like \texttt{SurvivalStan} \citep{brilleman2020} in Python have been developed. While they simplify the user-interface, they do not perform causal estimation. Similarly, Non-\texttt{Stan} based packages for Bayesian survival analysis such as \texttt{bayesSurvival}, the \texttt{BayesSurv} function in the \texttt{SemiCompRisks} package \citep{Kyu2014}, and \texttt{surv.bart} function in \texttt{BART} also do not provide causal effect estimates in standard point-treatment settings.

\texttt{causalBETA} (\texttt{B}ayesian \texttt{E}vent \texttt{T}ime \texttt{A}nalysis) addresses this gap by providing user-friendly functions for sampling from the posterior of semiparametric Bayesian event time models and performing g-computation to produce draws from the posterior distribution of causal contrasts. We provide the user with options for smoothing priors that can stabilize the model in sparse, small-sample settings. The functions have a familiar syntax identical to that in standard packages like \texttt{survival} and back-end to \texttt{Stan} for efficient posterior sampling using HMC. Custom \texttt{S3} classes and methods are provided so that familiar generic \texttt{R} functions such as \texttt{plot(...)} can be used. This improves user experience by removing the need to learn new functions. Moreover, the package contains detailed help files that can be accessed directly in \texttt{R}. In this paper we provide the methodological details of the implemented models, priors, and required causal assumptions. We demonstrate the package with an analysis of the \texttt{veteran} data set, which is publicly available in the \texttt{survival} package and can be accessed via the command \texttt{survival::veteran}.

\section{Data Structure, Causal Estimands, and Identification}

We consider a setting with $n$ independent subjects indexed by $i$. The observed data is of the form $D= \{ a_i, l_i, y_i, \delta_i \}_{i=1}^n$. We let $a_i\in\{0,1\}$ be a binary treatment indicator and $l_i \in\mathcal{L}$ be a set of $p$ covariates measured pre-treatment, which influence both assignment to one of the treatment groups as well as survival time. These may be a mix of discrete and continuous covariates. The observed event time, relative to time of treatment assignment, is given by $y_i =\min(t_i, c_i)$ and is the first of either an event at time $t_i$ or censoring at time $c_i$. The corresponding indicator of an event is $\delta_i = I(t_i <c_i)$. Throughout we use uppercase letters to denote random variables and lowercase letters to denote realizations.

The functions in \texttt{causalBETA} use the data $D$ to conduct posterior inference for causal estimands provided that key identification assumptions are met. In this section, we define these estimands formally in terms of potential outcomes and discuss the required identification assumptions. We let $T^a$ be a random variable representing the potential survival time for a subject had they been treated with treatment $A=a$. The functions in \texttt{causalBETA} provide posterior inference for population-level quantities of $T^a$ under different hypothetical treatments $a\in\{0,1\}$ as well as their contrasts. For some function $g$, these quantities take the form $E[g(T^a)]$, where expectation is with respect to the distribution of the potential outcome. Contrasts often take the form of differences, $E[g(T^1)] - E[g(T^0)]$, or ratios $E[g(T^1)]/E[g(T^0)]$. For example, if $g(T^a) = I(T^a>t)$, then $E[g(T^a)] = P(T^a >t)$ is the proportion of the target population that would have survived past time $t$ had everyone in the target population been, possibly counter to the fact, treated with $A=a$. Similarly, $\Psi(t) : = E[g(T^1)] - E[g(T^0)] = P(T^1 >t) - P(T^0 >t)$ contrasts the proportion surviving past time $t$ under treatment $A=1$ versus $A=0$. If this difference is zero, then there is no causal effect of treatment $A$ on the proportion surviving past $t$. 

The fundamental problem of causal inference is that we never observe survival rates in hypothetical worlds in which everyone in the target population receives each treatment $A=a$. Instead, some fraction of the target population is selected into treatment $A=1$ and another is selected into treatment $A=0$. In observational studies, this selection/treatment mechanism is not random and so neither the treated nor untreated groups are not representative of the target population. Instead selection is driven by factors, $L$, which also impact the outcome. For instance, patients with worse kidney function ($L$) may be more likely to get treatment $A=1$ and also more likely to die sooner. It is said this confounds the treatment effect since we cannot tease out whether the poor survival among those with treatment $A=1$ is due to the treatment or due to the fact that those getting the treatment had worse kidney function to begin with. Extrapolation of outcomes among those \textit{actually} treated/untreated to the target population can only be made under certain identification assumptions:

\begin{enumerate}
    \item Conditionally ignorable treatment mechanism: Conditional on observed covariates $L$, treatment is unrelated to potential outcomes
    $$ A \perp T^1, T^0 \mid L$$
    Said another way, a treated and untreated patients with the same features $L=l$ are comparable in terms of their potential outcomes. E.g. it should not the case that, even after adjusting for $L$, treated patients would have, say, been more likely to have shorter survival time even under no treatment. This assumption would be violated if there are unmeasred factors affecting both treamtent and outcome that are not included in $L$.
    \item Treatment Positivity: for each $l$ such that $f_L(l) >0 $,
    $$ P(A=1 \mid L=l) > 0$$
    This guarantees that both treatment options are possible for all subpopulations defined by $L$. If there is some $l$ such that $P(A=1 \mid L=l)=0$, then a treatment effect is not defined in this subgroup. This assumption can be violated either structurally (e.g. if $l$ is age and patients older than 65 are ineligible for treatment) or randomly due to small samples (e.g. treatment is possible at all ages, but in a particular data set there are very few patients older than 65 and they all happen to be untreated).
    \item Stable Unit Treatment Value Assumption (SUTVA): The observed outcome for a subject with treatment $A=a$ is equal to the potential outcome under that treatment.
    $$ T = AT^1 + (1-A)T^0 $$
    This assumption explicitly ties observed outcomes to potential outcomes and is often violated in settings with interference between subjects or where multiple verisons of the treatment exist.
\end{enumerate}

Assuming these assumptions hold, it is possible to identify each term of $\Psi(t)$ in terms of observed data distribution functions via a g-computation algorithm \citep{Robins1986,tsiatis2019}.
\begin{equation}
    \label{eq:gcomp}
    \begin{split}
        P(T^a > t) & = \int_{\mathcal{L}} P(T^a > t \mid L=l) dF_L(l)  \\ 
                   & = \int_{\mathcal{L}} P(T^a > t \mid A=a, L=l) dF_L(l)  \\ 
                   & = \int_{\mathcal{L}} P(T > t \mid A=a, L=l) dF_L(l)  \\ 
                   & = \int_{\mathcal{L}} \exp\Big( -\int_0^t \lambda(u \mid A=a, L=l) du \Big) dF_L(l) 
    \end{split}
\end{equation}

where $P(T > t \mid A=a, L=l)$ is the survival function for those treated with $A=a$ and having covariates $L=l$. The corresponding hazard function, $\lambda$, is defined as
$$ \lambda(t \mid A=a, L=l) = \lim_{dt\rightarrow 0 } \frac{ P( t \leq T <  t+dt \mid T\geq t, A=a, L=l) }{dt } $$
The first equality in \eqref{eq:gcomp} follows from averaging the conditional survival probability over the marginal distribution of $L$. In the second equality, ignorability is invoked to condition on a particular treatment value. Since positivity holds, each treatment is possible for each value of $L=l$, we can condition on $A=a$. The third line invokes SUTVA to replace the potential outcome with the observed outcome. This completes identification since the $P(T^a > t)$ is now expressed completely in terms of observed data. The final equality in \eqref{eq:gcomp} follows from the fact that for a given continuous random variable, $X$, the survival function $S(x)$ and hazard function $\lambda(x)$ have the relation $S(x) = \exp(-\int^x_0 \lambda(u)du)$.

When event times are right-censored, we can still conduct inference using contributions from both censored and failed subjects under two additional assumptions. First, censoring must be non-informative in the sense that $T \perp C \mid A, L$. This assumption is standard in survival analysis. Second, for each $t$ under consideration, there must be a positive probability of remaining uncensored up to that time, $P(C > t\mid A=a, L=l) > 0 $. A violation, $P(C > t\mid A=a, L=l) = 0$ for some $a, l$, suggests there is a subgroup of the target population who would always be censored by time $t$, making it impossible to learn about the survival distribution past that time. A practical implication of this is that one should avoid making inferences about $\Psi(t)$ at time points, $t$, greater than the maximum observed time since we have no data after that point.

According to Equation \eqref{eq:gcomp}, estimation of $P(T^a > t)$ can be done via a model for the hazard $ \lambda(t \mid A=a, L=l)$ and a model for the distribution of the confounders, $F_L(l)$. A posterior over $(\lambda,  F_L)$ induces a posterior over $\Psi(t)$. The required integrals are computed via Monte Carlo simulation. Accrodingly, the \texttt{causalBETA} package contains two key functions:\texttt{bayeshaz} outputs draws from the posterior of a user-specified hazard function and \texttt{bayesgcomp} takes the results of \texttt{bayeshaz} as input and performs Monte Carlo integration over the confounder distribution to obtain posterior draws of $P(T^a>t)$ and $\Psi(t)$.

\section{Bayesian Semiparametric Hazard and Confounder Models} \label{sc:model}

The \texttt{bayeshaz} function in the \texttt{causalBETA} package samples from the posterior of the following model for the hazard,
\begin{equation} \label{eq:hazmod}
    \lambda(t \mid A=a, L=l; \theta, \beta ) = \lambda_0(t; \theta ) \exp\Big( \beta(a, l) \Big)
\end{equation}
Above, $\lambda_0(t; \theta)$ is piecewise constant baseline hazard over a specified partition of the time axis, $ \lambda_0(t;\theta) = \sum_{k=1}^K I(\tau_{k-1} < t  \leq \tau_k ) \theta_k $, and $\beta(a, l)$ is a user-specified function, which we will call a ``regression function''. For compactness we denote the entire sequence of baseline hazards as $\theta = \{ \theta_k \}_{k=1}^K$. The package defaults to a partition with end points $\{ \tau_k \}_{k=0}^K$ such that $\tau_0:=0$ and $\tau_K :=\max_i y_i$. The end points are equally spaced so that $\tau_k - \tau_{k-1} = \max_i y_i / K$ for each $k$ and the number of intervals in the partition, $K$, is user-specified. Unlike the usual approach which leaves the baseline hazard unspecified and makes inference on $\beta(a,l)$ using a partial likelihood, proper Bayesian inference requires a fully specified likelihood. Moreover, the g-computation algorithm in Equation \ref{eq:gcomp} requires the full hazard, including the baseline hazard. Even though we must specify a model for the baseline hazard, this piecewise constant specification of the baseline hazard is effectively nonparametric. A large $K$, relative to the time scale of the observed data, yields a flexible model for the baseline hazard that can capture highly irregular observed hazard patterns.

The function $\beta(a,l)$ specifies how covariates impact the hazard. For example, one may specify $\beta(a, l) = \beta_1 a + l'\beta_2$. The \texttt{reg\_formula} argument of \texttt{bayeshaz} allows the user to specify the function $\beta(a, l)$ using familiar survival regression syntax, e.g.
\begin{lstlisting}
    reg_formula = Surv(y, delta) ~ A + L
\end{lstlisting}
More examples and details will be given in Section \ref{sc:demo}. In the following we will abuse notation slightly and let $\beta$ denote the collection of coefficients parameterizing $\beta(a, l)$, e.g. $\beta=(\beta_1, \beta_2)$ in the previous example. Here, treatment and covariates impact the hazard via a scaling of the baseline hazard so that $\exp(\beta_1)$ and $\exp(\beta_2)$ have the usual hazard ratio interpretations. In this specification, for instance, the hazard among treated patients with covariates $l$ is $\exp(\beta_1)$ times the hazard among untreated patients with the same covariates. If $\exp(\beta_1) > 1$, then the hazard is larger among the treated. Note however that contrasts of hazard rates are not, in general, causally meaningful \citep{hernan2010}. This is because the hazard at time $t$ is by definition conditional on survival up to time $t$, $T > t$, which is a post-treatment event. On the other hand, contrasts of survival probability such as $\Psi(t)$ are valid since they are marginal probabilities rather than probabilities conditional on post-treatment events. From a causal perspective, $\beta_1$ and $\beta_2$ are only nuisance parameters needed to plug into the g-formula in \eqref{eq:gcomp} and compute $\Psi(t)$.

In addition to a model for the event hazard, we also need a model for the confounder distribution. In Bayesian causal inference, a canonical nonparametric choice is the Bayesian bootstrap \citep{Oganisian2021}, which places probability mass on each unique confounder vector observed in the sample,
$$F_L(l) = \sum_{i=1}^n \pi_i I( l_i \leq l) $$
Here, $\{ \pi_i \}_{i=1}^n$ is a sequence of non-negative weights that sum to 1. Note the empirical distribution is the case with uniform weights $\pi_i = 1/n$ for each $i$. Instead of treating the weights as fixed, the Bayesian bootstrap treats them as unknown with an improper Dirichlet prior $(\pi_1, \pi_2, \dots, \pi_n) \sim Dir( 0_n)$. Here, $0_n$ is the length-$n$ zero vector.

\subsection{Prior Processes for the Baseline Hazard} \label{sc:prior_process}
When running the model in \eqref{eq:hazmod}, \texttt{bayeshaz} specifies priors on the parameters governing the hazard function, $\theta=\{ \theta_k \}_{k=1}^K$ and $\beta$. The default prior for the length $p$ vector of coefficients, $\beta$, is multivariate normal with mean $0_p$ and covariance matrix $\sigma^2 I_p$. Here, $0_p$ and $I_p$ denote the length $p$ zero-vector and $p\times p$ identity matrix, respectively. The hyperparameter $\sigma$ can be set by the user using the \texttt{sigma} argument in \texttt{bayeshaz}. The default value is $\sigma=3$. This is uninformative on the hazard ratio scale as it puts 95\% prior probability on hazard ratios in the range $[0.003, 357]$. Setting \texttt{sigma} to larger values is not recommended.

For the baseline hazard parameters, $\theta$, \texttt{bayeshaz} uses a prior process that smooths the hazards across the $K$ intervals. Having such a smoothing prior is crucial because, all else being equal, we would prefer a fine partition with a large $K$. However, a large $K$ means that very few events may occur in any one interval. This makes unregularized estimation erratic, especially in later time intervals that may have fewer patients at-risk. By default, \texttt{bayeshaz} sets $K=100$, but other values can be specified via the argument \texttt{num\_partition}. Across these increments, \texttt{bayeshaz} places a first-order autoregressive prior on the log baseline hazard rates. Specifically, let  $\tilde \theta_k = \log \theta_k$ be the log baseline hazard rate at interval $k$. The process is defined by the initial condition $\tilde \theta_1 = \eta + \nu_1 \epsilon_1$ and, for $k=2,3,\dots, K$,
$$ \tilde \theta_k = \eta ( 1 - \rho) + \rho \tilde \theta_{k-1} + \nu_k \epsilon_k$$
where, $\epsilon_k \stackrel{iid}{\sim} N(0,1)$ are noise terms and $-\infty <\eta < \infty$, $-1<\rho<1$, and $\nu_k>0$ are the parameters of the process. This prior process has a constant mean $E[\tilde \theta_k] = \eta$ and variance $V[\tilde \theta_k] = \nu_k^2 / (1-\rho^2)$. Importantly, however, the log-hazard at each interval is allowed to depend on the log-hazard in the previous interval. This means that the correlation between the log hazard at some interval $v$ and the log-hazard $u$ intervals prior is $Corr(\tilde \theta_v, \tilde \theta_{v-u}) = \rho^u$. Since $0<\rho<1$, correlation is larger with more recent intervals and smaller with earlier ones. This encodes the prior believe that, in the absence of data in interval $v$, the log-hazard rate in this interval should be similar to log-hazard rates in recent intervals - thus smoothing the baseline hazard rate across time. The \texttt{bayeshaz} function provides users with two versions of this prior. This first sets $\rho=0$, which eliminates smoothing across time and yields independent estimation of each $\tilde \theta_k$ - each shrunk towards the overall prior mean log hazard rate, $\eta$. The second places a $Beta(2,2)$ prior on $\rho$, where $Beta(2,2)$ is the four parameter Beta distribution with the two shape parameters set to 2. The other two parameters are the minimum and maximum, set to -1 and 1 respectively to provide support in the region $-1 \leq \rho \leq 1$. Note the usual two-parameter Beta is a special case with minimum and maximum set to 0 and 1, respectively, and having support in the interval $[0,1]$. The prior is centered a mean of $E[\rho] = 0$ (no correlation across time) but with 95\% prior probability on values in $-.8<\rho<.8$ - thus allowing for correlation in either direction \textit{a priori}. The corresponding prior density implied by the autoregressive process is 
\begin{equation}
    f(\{ \tilde \theta_k \}_{k=1}^K; \rho, \eta, \{\nu_k\}_{k=1}^K ) = N( \tilde \theta_1; \eta, \nu_1^2) \prod_{k=2}^K N( \tilde \theta_k; \eta(1-\rho) + \rho \tilde \theta_{k-1}, \nu_k^2  )
\end{equation}
where $N(x;\mu, \sigma^2)$ denotes the normal density with mean $\mu$ and variance $\sigma^2$ evaluated at $x$. For the remaining hyperparameters, \texttt{bayeshaz} defaults to $\eta\sim N(0,1)$ and $\nu \stackrel{iid}{\sim} Gam(1,1)$. The former implies a log-normal hyperprior on the prior mean hazard rate, $\exp(\eta)$, which has a long tail that allows for large average hazards. The latter is a standard non-informative prior on the standard deviations. Thus, the full prior density is given by 
\begin{equation} \label{eq:prior}
    N_p(\beta; 0_p, \sigma^2 I_p)\cdot f(\{ \tilde \theta_k \}_{k=1}^K; \rho, \eta, \{\nu_k\}_{k=1}^K ) \cdot N(\eta; 0,1) \cdot Beta(\rho; 2,2) \cdot  \prod_{k=1}^K Gam(\nu_k; 1,1) 
\end{equation}
Where $N_p(\beta; 0_p, \sigma^2 I_p)$, $N(\eta; 0,1)$, $Beta(\rho; 2,2)$, and $Gam(\nu_k; 1,1)$ represent the density functions of the associated prior distributions.

\section{Posterior Sampling and G-Computation} \label{sc:computation}

Given the data $D$, the likelihood under the model in \eqref{eq:hazmod} is given by 
$$ \mathcal{L}( D ; \theta, \beta ) = \prod_{i=1}^n \exp\Big(-\int_{0}^{y_i}  \lambda(u \mid a_i, l_i; \theta, \beta )du \Big )  \lambda(y_i \mid a_i, l_i; \theta, \beta )^{\delta_i} $$
It has been shown that this likelihood can be expressed in terms of a Poisson likelihood \citep{Laird1981}, which makes specification in \texttt{Stan} quite simple. We provide details of this in Appendix A. The likelihood above combined with the prior in \eqref{eq:prior} are automatically specified in \texttt{Stan} when the user calls \texttt{bayeshaz}. \texttt{Stan} then runs a variant of HMC \citep{neal2011mcmc} to obtain draws of the hazard parameters. We obtain draws of the confounder distribution parameters, $\{ \pi_i \}_{i=1}^n$ via conjugacy. Under a $Dir(0_n)$ prior, the posterior is $\{ \pi_i \}_{i=1}^n \sim Dir(1_n)$, where $1_n$ is length-$n$ vector of ones. 

The function \texttt{bayeshaz} returns $m=1,2,\dots, M$ draws from the posterior of all these unknown parameters,  $ \Big( \{\pi_i^{(m)} \}_{i=1}^n, \{ \theta_k^{(m)} \}_{k=1}^K, \beta^{(m)} \Big)$. The function \texttt{bayesgcomp} takes these draws as input and does the requisite post-process to obtain draws of the causal parameters of interest. This necessarily involves a Monte Carlo approximation of the integral in \eqref{eq:gcomp}. Specifically, for the $m^{th}$ posterior draw \texttt{bayesgcomp} computes a draw of $\Psi(t)$ for a range of specified $t$ as follows. Under intervention $A=a$,
\begin{enumerate}
    \item For each observed $l_i$ in the sample, simulate $B$ survival times using a posterior draw of the hazard, 
    $$ T^{(m), (1)}, T^{(m), (2)}, \dots, T^{(m), (B)}  \sim \lambda( t \mid A=a, L=l_i; \beta^{(m)}, \{ \theta_k^{(m)} \}_{k=1}^K )$$
    this is done using the function \texttt{mets::rchaz}, a computationally efficient implementation of the inverse-CDF method.
    \item After having done the above for each subject $i$, we approximate the conditional survival probability by averaging across the $B$ simulations
    $$ P^{(m)}(T > t \mid A=a, L=l_i) \approx \frac{1}{B} \sum_{b=1}^B I(T^{(m), (b)} > t ) $$
    At each iteration $m$ and for each subject $i$, the above is done for all time points $t$ under consideration. This ensures monotonicity of the conditional survival curve within each posterior draw.
    \item We do steps 1-2 under each treatment $a\in\{0, 1\}$. Finally, we obtain a draw of the marginal survival rate difference by averaging over the Bayesian bootstrap model for $F_L$ given $\{\pi_i^{(m)} \}_{i=1}^n$, 
    \begin{equation*}
        \begin{split}
            \Psi^{(m)}(t) & = \int_{\mathcal{L}} \Big\{ P^{(m)}( T >t \mid A=1, L=l) - P^{(m)}( T >t \mid A=0, L=l) \Big\}dF_L(l)  \\
                & \approx \sum_{i=1}^n \pi_i^{(m)}  \Big\{  P^{(m)}( T >t \mid A=1, L=l_i) - P^{(m)}( T >t \mid A=0, L=l_i) \Big\} \\
        \end{split}
    \end{equation*}
\end{enumerate}

Repeating the above for each posterior draw yields a set of draws from the posterior survival curve under each intervention $a\in\{0,1\}$, $\{ P^{(m)}(T^a > t) \}_{m=1}^M$, as well as the causal contrasts, $\{ \Psi^{(m)}(t)\}_{m=1}^M$. The package takes the mean across the $M$ draws to form posterior point estimates and takes the $(\alpha/2)$th and $(1-\alpha/2)$th percentiles of the $M$ draws to form a $(1-\alpha)\%$ credible interval, for a given credible level $\alpha \in (0,1)$. The number of Monte Carlo simulations, $B$, can be specified using the \texttt{B} argument in \texttt{bayesgcomp}. By default, \texttt{bayesgcomp} does the above at each $t$ equal to the midpoint of the $K$ intervals. However, users may specify a vector of arbitrary time points of interest instead.

\section{Demonstration: Efficacy of Chemotherapy Treatments on Survival Among Patients with Lung Cancer}
\label{sc:demo}

In this section we will illustrate the use of \texttt{causalBETA} by analyzing a lung cancer data set publicly available in the \texttt{survival} package and accessed in \texttt{R} via \texttt{survival::veteran}, and the code for demonstration is available under the GitHub repository of \texttt{causalBETA} \footnote{\url{https://github.com/RuBBiT-hj/causalBETA/tree/main/demo_code}}. These data, published by \cite{kalbfleisch2011}, are from a US Veterans Administration study of males with inoperable lung cancer. In the study, $n=137$ patients were given either a standard course of therapy or a novel test therapy. In the data, the former is coded as \texttt{trt=1} and the latter is coded as \texttt{trt=2}, which we recode as \texttt{A=0} and \texttt{A=1} to connect with standard causal notation used in this paper. Although the data are from a randomized trial, the public availability of these data are ideal for demonstration purposes. In the data set, the observed time is measured in days and given by \texttt{time}. The death event indicator is given by \texttt{status}, with \texttt{status=1} indicating death and \texttt{status=0} indicating censoring. Similarly, we rename these to \texttt{y} and \texttt{delta}, respectively, to facilitate connections with notation used earlier in the manuscript. There were 68 patients randomized to $A=1$ and 69 patients to $A=0$. For each patient, the data set contains a factor column called \texttt{celltype}, which indicates which of four cell types each patient's tumors are comprised of (\texttt{squamous}, \texttt{smallcell}, \texttt{adeno}, or \texttt{large}). Since the functions in \texttt{causalBETA} require one-hot encoding of factor variables, this column is converted to four 0/1 indicators - one for each type. Other covariates are available in the dataset. A full description can be found by accessing the help file using \texttt{?survival::veteran}. Below, we install \texttt{causalBETA} from the GitHub repository, load it along with other packages, and perform the data manipulation just discussed. Note that \texttt{causalBETA} is build on top of the \texttt{cmdstanr} interface to \texttt{Stan}. At the time of drafting this manuscript, installation of \texttt{cmdstanr} version 0.5.3 or higher is therefore required along with a \texttt{C++} toolchain. Installation instructions and resources can be found here: \url{https://github.com/stan-dev/cmdstanr}.

\begin{lstlisting}
## install and load causalBETA
devtools::install_github("RuBBiT-hj/causalBETA")
library(causalBETA)

## load other packages
library(survival)

## load data and re-code variables
data = survival::veteran
data$A = 1*(data$trt==2)

## rename variables
var_names = colnames(data)
colnames(data)[var_names=='status'] = 'delta'
colnames(data)[var_names=='time'] = 'y'

## append one-hot encoded celltypes
data = cbind(data, model.matrix(data=data, ~ -1 + celltype))
\end{lstlisting}

Our goal in this demonstration will be to estimate the causal effect of the novel test therapy ($A=1$) versus standard therapy ($A=0$) on survival probability. The primary efficacy measure will by $\Psi(t) = P(T^1 > t) - P(T^0 >t)$ for $t$ within the support of the data. This represents the difference in the proportion of patients with inoperable lung cancer who would have survived past day $t$ had they all received the novel test chemotherapy versus standard chemotherapy. In these data, only nine subjects were censored and the maximum observed time was 999 days. Thus, inferences about $\Psi(t)$ can only be made reliably up to at most $t=999$. We will first perform and unadjusted analysis and then a confounder-adjusted analysis. Though adjustment is not strictly necessary due to randomization, we will do it to illustrate syntax.

\begin{figure}[h!]
    \centering
    \includegraphics[width=1\linewidth]{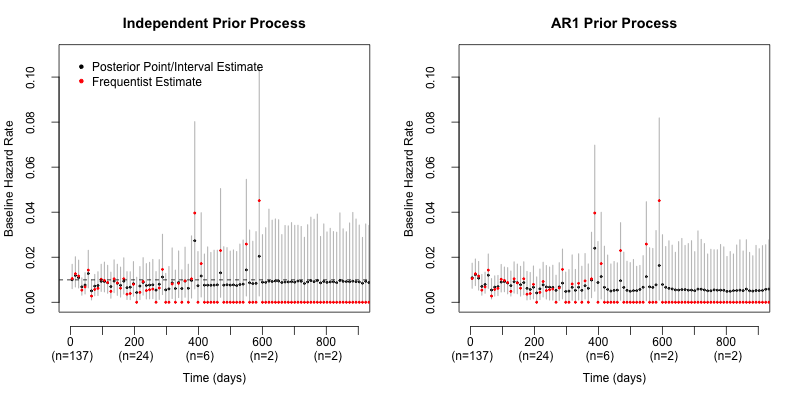}
    \caption{Posterior distribution over the baseline hazard function from an unadjusted model (i.e. hazard of death among patients with $A=0$). Left panel: model under independent prior process over the hazard rates (\texttt{model='indepndent'} in \texttt{bayeshaz}). Right panel: model under AR1 prior process over the hazard rates (\texttt{model='AR1'}). Black point/segments represent posterior mean and 95\% credible interval for the hazard at each of the \texttt{num\_partitions=100} intervals. At-risk counts are given along the x-axis. Frequentist point estimates from \texttt{eha::pchreg} are overlaid as red points. The dashed black line in the left panel is the overall average hazard rate across intervals}
    \label{fig:unadjhazplot} 
\end{figure}

\subsection{ An Unadjusted Analysis: Introduction to Key Functions}

To perform an unadjusted analysis, we first estimate the hazard model in \eqref{eq:hazmod} with no covariates, $l$, present so that $\beta(a,l) = \beta_1 a$. This leads to the hazard model $\lambda(t \mid A=a) = \lambda_0(t)\exp(\beta_1a)$. The baseline hazard, $\lambda_0(t)$ is given the piecewise constant form in \eqref{eq:hazmod}. The function \texttt{bayeshaz} in \texttt{causalBETA} can be used to draw from the posterior of the unknown baseline hazard and $\beta_1$ as follows:

\begin{lstlisting}
set.seed(1) ## set seed so MCMC draws are reproducible

post_draws_ind = bayeshaz(d = data, ## data set
                          reg_formula = Surv(y, delta) ~ A, 
                          num_partitions = 100, 
                          model = 'independent',
                          sigma = 3, 
                          A = 'A',
                          warmup = 1000, 
                          post_iter = 1000) 
\end{lstlisting}

The \texttt{reg\_formula} argument specifies the form of $\beta(a,l)$ in Equation \eqref{eq:hazmod} and here it is set to only model the hazard as a function of the treatment indicator. The \texttt{Surv(y,delta)} syntax is identical to the syntax used in the regression functions in the \texttt{survival} package, where the first argument specifies the observed time and the second argument is an indicator of an event (with a one) and censoring (with a zero). By default, the package uses $K=100$ partitions to estimate the baseline hazard, but the user can specify other values via the \texttt{num\_partitions} argument above. The argument \texttt{model = 'independent'} corresponds to the autoregressive prior described in Section \ref{sc:prior_process} with $ \rho = 0$. That is, the hazards over the increments are \textit{a priori} independent with separate variances, $\nu_k^2$, but centered around a common mean hazard, $\exp(\eta)$. The arguments \texttt{warmup=1000} and \texttt{post\_iter=1000} specify the number of warm-up (sometimes called burn-in) draws and the number of post-warm-up draws from the posterior to retain. Thus, \texttt{post\_iter} is what we call $M$ in this paper. The function call above will return \texttt{post\_iter=1000} posterior draws of all model unknowns. The arguments \texttt{A=} specifies the column name corresponding to the binary treatment indicator in the data frame specified by \texttt{d=}. All variables specified in \texttt{reg\_formula} should be present in \texttt{d}. Finally, the argument \texttt{sigma=} is the prior standard deviation, $\sigma$, on the independent mean-zero Gaussian priors on each of the coefficients in the hazard model. The default is $\sigma=3$.

The baseline hazard function from this model is presented on the left panel of Figure \ref{fig:unadjhazplot}. Notice that the independent prior process shrinks towards a constant baseline hazard prior at $\exp(\eta)$ (visualized by the dashed black line). In the later time intervals (t>400) there are very few patients at risk and so the posterior estimates are shrunk towards the prior constant-hazard rate of $\exp(\eta)$. The credible intervals get larger, reflecting the sparsity of the data over these time points. 

The right panel of Figure \ref{fig:unadjhazplot} presents alternative estimates under the autoregressive prior process for the baseline hazard which is obtained by running \texttt{bayeshaz} with  \texttt{model = 'AR1'}. Notice that the estimates are visibly auto-correlated and so, unlike the independent prior process, exhibit a wavy trajectory. In the absence of data, the posterior of the hazard at each interval is pulled towards a compromise of the overall mean of the process $\eta$ and the estimate from the previous interval. Borrowing information across intervals leads to narrower intervals than under the independent prior process.

Notice how erratic the frequentist estimates are in Figure \ref{fig:unadjhazplot} - oscillating between zero and very large values whenever an event occurs. Of course, we do not actually think there is no possibility of death at, say, day 800 among those alive at that time. Rather, the zero estimate is just due to small sample variability. Similarly, the spikes in the hazard function around day 400 and 600 are due to the low at-risk count. The Bayesian model under either prior, however, smooths out these frequentist estimates.

The plots in Figure \ref{fig:unadjhazplot} are easily produced using the base \texttt{plot} function in \texttt{R}. For instance, the plot on the left panel was produced using:

\begin{lstlisting}
plot(post_draws_ind, 
     ylim=c(0,.11), 
     xlim=c(0, 900),
     type='p', 
     main='Independent Prior Process', 
     ylab = 'Baseline Hazard Rate', 
     xlab = 'Time (days)')

legend(x=0, y=.111, 
       legend = c('Posterior Point/Interval Estimate', 
                  'Frequentist Estimate'), 
       col=c('black', 'red'), pch=c(20,20), bty='n')
\end{lstlisting}

This is possible because \texttt{causalBETA} contains custom \texttt{S3} classes for objects produced by its functions. Corresponding methods for base \texttt{R} generic functions such as \texttt{plot} are also included. For example, the object \texttt{post\_draws\_ind} is an object of class \texttt{bayeshaz}. Therefore, when the generic \texttt{plot} function is called above, the method \texttt{plot.bayeshaz} is invoked under the hood. The advantage is that no new plotting functions are introduced to the user, while access to familiar arguments such as \texttt{ylim}, \texttt{xlab}, etc. are retained. As can be seen, the usual functions such as \texttt{legend} can be used to add layers to the initial plot.

Moreover, the object \texttt{post\_draws\_ind} contains the full set of posterior draws for all unknowns in the model which immediately accessible to the user. For instance, the object \texttt{post\_draws\_ind\$haz\_draws} is a $1000 \times 100$ (generally, \texttt{post\_draws} $\times$ \texttt{num\_partitions}) matrix of posterior draws - 1000 draws for each of the 100 hazard rates in the partition. Thus, to add the average of the posterior mean hazard rates (the dashed horizontal line on the left panel of Figure \ref{fig:unadjhazplot}) we simply write the following code after running \texttt{plot} in the previous code block:

\begin{lstlisting}
abline(h=mean(colMeans(post_draws_ind$haz_draws)), lty=2, col='black')
\end{lstlisting}

Similarly, \texttt{post\_draws\_ind\$beta\_draws} is a $1000 \times 1$ matrix of posterior draws for $\beta_1$. In general, there will be one column for each coefficient in $\beta(a,l)$. Access to the underlying draws are useful if the analyst wishes to make custom transformations of the draws and also to do convergence checking, which we will discuss in another section. In addition to the posterior draws, various meta-data about the \texttt{bayeshaz} function call are stored in \texttt{post\_draws\_ind}. For example \texttt{post\_draws\_ind\$midpoints} is a vector containing the midpoints of the 100 partition intervals specified in the previous \texttt{bayeshaz} call using \texttt{num\_partition=100}. Similarly, \texttt{post\_draws\_ind\$partition} contains the length \texttt{num\_partition+1} vector of partition endpoints. These can be useful for plotting or constructing other descriptive statistics. For instance, the frequentist estimates shown in red in Figure \ref{fig:unadjhazplot} were obtained and plotted by running the following code after the \texttt{plot(post\_draws\_ind,...)} call shown previously:
\begin{lstlisting}
## pchreg in eha package estimates piecewise constant hazard regression
freq_res = eha::pchreg(data=data, 
                       ## specify endpoints of intervals
                       cuts = post_draws_ind$partition, 
                       formula = Surv(y, delta) ~  A) ## hazard formula

## add points to plot
points(post_draws_ind$midpoint, freq_res$hazards, col='red',pch=20, cex=.5) 
\end{lstlisting}
The \texttt{pchreg} function in the package \texttt{eha} computes estimates of the piecewise constant baseline hazard model in Equation \ref{eq:hazmod} using maximum-likelihood estimation. Such plots can also serve as posterior-predictive checks, as survival times may be simulated from the posterior predictive and plotted against thee empirical distribution to assess model fit.

\subsection{Covariate-Adjusted Analysis and G-Computation}

In observational studies, we often want to adjust for covariates which we believe influence both selection into a treatment group and the outcome. Covariates can be added into the hazard model via specification of the \texttt{reg\_formula} argument in the \texttt{bayeshaz} function as follows:
\begin{lstlisting}
formula1 = Surv(y, delta) ~ A + age + karno + celltypesquamous + celltypesmallcell + celltypeadeno

post_draws_ar1_adj = bayeshaz(d = data, 
                              reg_formula = formula1,
                              model = 'AR1',
                              A = 'A', 
                              warmup = 1000, post_iter = 1000) 
\end{lstlisting}

\begin{figure}[h!]
    \centering
    \includegraphics[width=1\linewidth]{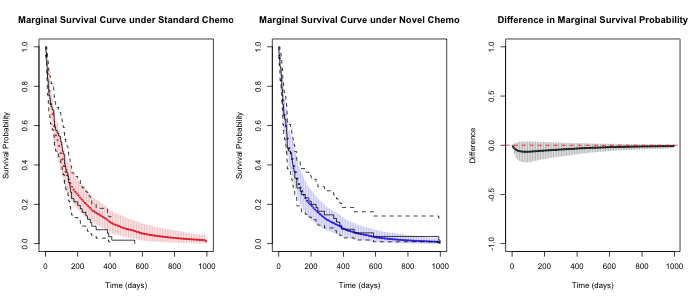}
    \caption{Posterior distribution over the marginal survival function, $P(T^a>t)$, for standard chemo ($a=0)$ in the left panel, for novel chemo ($a=1$) in the middle panel, and the difference $\Psi(t) = P(T^1>t) - P(T^0>t)$ in the right panel. Posterior esitmates are based on covariate-adjusted hazard model. At each time $t$, points represent posterior mean and segments represent 95\% credible intervals. For comparison, the black lines and on the first two panels represent the Kaplan-Meier (KM) estimates computed using \texttt{survival::survfit} function. We see that covariate adjustment does not move the survival curves much from their unadjusted KM estimates. This is perhaps because the treatment was randomized and, therefore, not confounded. The third panel indicates very little difference in efficacy across time. }
    \label{fig:adjsurvplot}
\end{figure}

Above, we include three tumor cell type indicators for the four possible cell type levels. We also include age and Karnofsky performance score (\texttt{karno}) (see here for more details: \url{https://www.hiv.va.gov/provider/tools/karnofsky-performance-scale.asp}. Note that we are fitting a model with an AR1 prior process on the baseline hazard, as indicated by \texttt{model='AR1'}. Interactions can be specified in the usual way after the \texttt{\~} in the formula. For instance, an interaction between treatment and age can be specified as 
\begin{lstlisting}
Surv(y, delta) ~ A + age + A*age + karno + celltypesquamous + celltypesmallcell + celltypeadeno
\end{lstlisting}
Equivalently, the following syntax will automatically include main effects and the interaction effect:
\begin{lstlisting}
Surv(y, delta) ~ A*age + karno + celltypesquamous + celltypesmallcell + celltypeadeno
\end{lstlisting}
As in base \texttt{R}, the following syntax will include main effects of treatment and all covariates, as well as an interaction effect between treatment and each covariate:
\begin{lstlisting}
Surv(y, delta) ~ A*(age + karno + celltypesquamous + celltypesmallcell + celltypeadeno)
\end{lstlisting}
We fit the main effects model specified in \texttt{formula1} for illustration. Just as before, we have access to posterior draws of all model coefficients which are stored in \texttt{post\_draws\_ar1\_adj\$beta\_draws}. By construction, this is an object of class \texttt{mcmc}, making it fully compatible with methods for generic functions contained in the \texttt{coda} package, a popular \texttt{R} package for summarizing MCMC draws. For instance, the posterior of the beta coefficients can be summarized using the generic \texttt{summary} function as follows:
\begin{lstlisting}
    summary(post_draws_ar1_adj$beta_draws, quantiles = c(.025, .975))
\end{lstlisting}
The \texttt{quantiles} argument specified the 2.5th and 97.5 percentiles of the posterior to be printed for each coefficients. These can serve as 95\% credible intervals. The output is:
\begin{lstlisting}
> summary(post_draws_ar1_adj$beta_draws, quantiles = c(.025, .975))

Iterations = 1:1000
Thinning interval = 1 
Number of chains = 1 
Sample size per chain = 1000 

1. Empirical mean and standard deviation for each variable,
   plus standard error of the mean:

                      Mean       SD  Naive SE Time-series SE
A                  0.24801 0.187073 0.0059158      0.0076333
age               -0.01609 0.008288 0.0002621      0.0003941
karno             -0.03767 0.004806 0.0001520      0.0001745
celltypesquamous  -0.55906 0.269659 0.0085274      0.0118384
celltypesmallcell  0.34599 0.259894 0.0082186      0.0130015
celltypeadeno      0.68840 0.281906 0.0089146      0.0113417

2. Quantiles for each variable:

                      2.5%      97.5%cnn
A                 -0.12184  0.6066283
age               -0.03245 -0.0004157
karno             -0.04718 -0.0283638
celltypesquamous  -1.07121 -0.0046759
celltypesmallcell -0.16084  0.8858032
celltypeadeno      0.12949  1.2571858
\end{lstlisting}
In the top panel of printed results, the Mean and SD columns print the mean and standard deviation of the posterior draws. More information about the \texttt{coda} summary function can be found by accessing the help file via \texttt{?summary.mcmc }. This output may be useful for associational analysis but, as discussed in Section \ref{sc:model}, the posterior mean coefficient of \texttt{A}, computed above to be \texttt{0.24801}, does not have a causal interpretation. Instead, we use a second function, \texttt{bayesgcomp}, which takes as input a set of posterior draws from \texttt{bayeshaz}, and performs g-computation to obtain causal contrasts. This can be done as follows:

\begin{lstlisting}
gcomp_res = bayesgcomp( post_draws_ar1_adj, ## bayeshaz output
                    ref = 0, ## treatment reference group
                    B = 1000) ## monte carlo iterations in g-comp
\end{lstlisting}

Above, the argument \texttt{B} corresponding to the number of Monte Carlo iterations used to approximate the integral over the confounder distribution, denoted as $B$ in Section \ref{sc:computation}. The object \texttt{gcomp\_res} contains posterior draws of the marginal survival curves under a reference treatment (which we specify to be $A=0$ via \texttt{ref=0} in the function call) and the other treatment (which is of interest). These can be accessed via \texttt{gcomp\_res\$surv\_ref} and \texttt{gcomp\_res\$surv\_trt} respectively. By default, \texttt{bayesgcomp} computes marginal survival curves at each of the midpoints of the \texttt{num\_partitions} intervals specified in \texttt{bayeshaz}. Both \texttt{gcomp\_res\$surv\_ref} and \texttt{gcomp\_res\$surv\_trt} will by matrices with dimension \texttt{post\_iter} $\times$ \texttt{num\_partitions}. Additionally, \texttt{gcomp\_res\$ATE} is a matrix with the same dimension that contains posterior draws of the survival difference, $\Psi(t)$, for each $t$ that is a midpoint in the partition. The function \texttt{bayesgcomp} also has an arguement \texttt{t}, which can be a vector of user-specified time values at which the marginal survival functions and their contrast, $\Psi(t)$, can be computed. This is useful as many analyses in biomedical fields are only concerned with, say, contrasting differences in 2-year, 3-year, and 5-year survival probabilities. Figure \ref{fig:adjsurvplot} plots the resulting marginal survival curves and $\Psi(t)$. 

Just as with \texttt{bayeshaz}, custom \texttt{S3} methods for \texttt{R} generic functions such as \texttt{plot} are available and can be used to easily generated figures such as the one in Figure \ref{fig:adjsurvplot}. For example, the left panel of Figure \ref{fig:adjsurvplot} was generated with the following code:
\begin{lstlisting}
 plot(gcomp_res, mode = 0, type ='p', 
     ylim=c(0,1), xlim=c(0, 1000), 
     xlab='Time (days)', ylab='Survival Probability',
     main='Marginal Survival Curve under Standard Chemo')
## overlay kaplan-meier
d0 = data[data$A==0, ]
lines(survfit(data=d0, Surv(y, delta)~ 1))   
\end{lstlisting}

The argument \texttt{mode=0} results in the first panel - the survival curve for the reference treatment group, specified to be \texttt{ref=0} in the \texttt{bayesgcomp} function call. Specifying \texttt{mode=1} results in the second panel in Figure \ref{fig:adjsurvplot} for the non-reference group. The third panel is obtained via \texttt{mode='ATE'}. Specifying \texttt{mode=c(0,1)} plots the marginal survival curves for both treatment groups on the same plot. Additional arguments are described in the help file accessed via \texttt{?plot.ATE}. For example, by default the figures plot 95\% credible intervals, but other credible probability levels such as 90\% intervals may be specified by \texttt{level\_CI=.90}. Options such as colors and labels are available as well.

\subsection{Assessing Posterior Convergence and Monte Carlo Error}

Bayesian causal inference is done using the posterior draws of the causal quantities of interest, obtained via the computational procedures described in Section \ref{sc:computation}. Thus, inference depends on whether the underling MCMC chains for these causal parameters have converged. Many methods for assessing convergence have been proposed and implemented in the \texttt{coda} package in \texttt{R}. Since the function \texttt{bayesgcomp} outputs the relevant draws as a matrix of class \texttt{mcmc}, the functions in \texttt{coda} can be used directly on these draws to assess convergence. We will not go over the details of these checks which are already discussed elsewhere \citep{Brookes1998, gelman2013} but will illustrate implementation in the \texttt{causalBETA} packages.

Specifically, we will consider estimateion of $\Psi(t)$ for $t\in\{ 365,2\cdot 365 \}$ - i.e. difference in the proportion of the population surviving past 1 and 2 years under treatment $A=1$ vs. $A=0$. To assess convergence it is often helpful to re-run the procedure in Section \ref{sc:computation} a few times (say, three times) with three different starting values. This yields three sets of draws, $\{ \Psi^{(m)}(t) \}_{m=1}^M$, at each $t$. These are referred to as ``chains.'' We can do this in our package by calling \texttt{bayeshaz} and \texttt{bayesgcomp} in a loop for three iterations as follows. 

We first generate and empty list of mcmc draws using the \texttt{coda::mcmc.list} function, \texttt{ATE\_chains = coda::mcmc.list()}. We then run the MCMC algorithm to obtain $M=1000$ draws after a 2000 draw warmup using \texttt{bayeshaz}. After obtaining these, we run g-computation and obtain a set of draws $\{ \Psi^{(m)}(t) \}_{m=1}^M $ using \texttt{bayesgcomp}. We store the results as an element in the list \texttt{ATE\_chains}.

\begin{lstlisting}
## run three chains:
n_chains = 3

## create an empty list which will contain the three chains/sets of draws
ATE_chains = coda::mcmc.list()

set.seed(1)
for(chain in 1:n_chains){
        ## obtain posterior draw of survival distribution parameters
        post_draws = bayeshaz(d = data, 
                              reg_formula = formula1,
                              model = 'AR1',
                              A = 'A', 
                              warmup = 2000, 
                              post_iter = 1000) ## output 1000 draws
        
        ## run g-computation to compute 1yr and 2yr survival rate difference
        gcomp_res = bayesgcomp(post_draws_ar1_adj , ref = 0, t = c(365, 2*365) )
        
        ## store posterior draws (1000 by 2) matrix
        ATE_chains[[chain]] = gcomp_res$ATE
}
\end{lstlisting}
Note the specification of \texttt{t = c(365, 2*365)} instructs \texttt{bayesgcomp} to evaluate the contrast at only the specified time values. We can easily compute posterior summaries of these contrasts using the combined set of draws:
\begin{lstlisting}
> summary(ATE_chains, quantiles = c(.025, .975))

Iterations = 1:1000
Thinning interval = 1 
Number of chains = 3 
Sample size per chain = 1000 

1. Empirical mean and standard deviation for each variable,
   plus standard error of the mean:

         Mean      SD  Naive SE Time-series SE
[1,] -0.03894 0.03060 0.0005588      0.0004702
[2,] -0.01532 0.01352 0.0002469      0.0002124

2. Quantiles for each variable:

         2.5%    97.5%
[1,] -0.09904 0.018145
[2,] -0.04579 0.006693
\end{lstlisting}
For instance, above we see that the posterior mean of $\Psi(365)$ is \texttt{-0.03894} with a 95\% credible interval \texttt{[- 0.09904, 0.018145]}. For $\Psi(2\cdot 365)$ we have \texttt{-0.01532} with interval \texttt{[- 0.04579, 0.006693]}.

\begin{figure}[h!]
    \centering
    \includegraphics[width=1\linewidth]{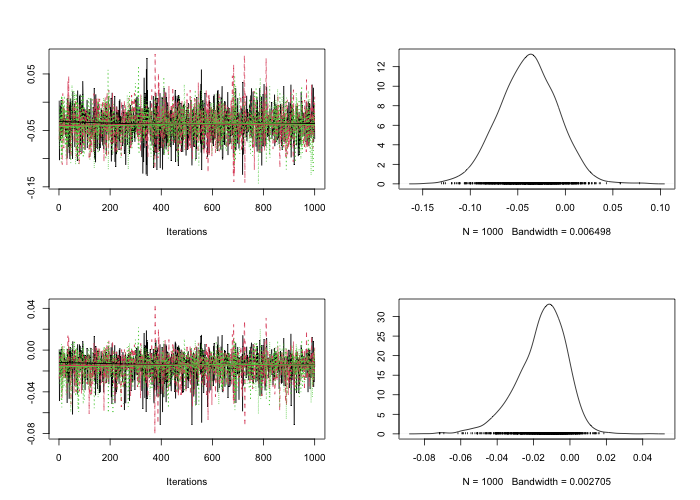}
    \caption{First column: Traceplots of draws $\{\Psi(t)\}_{m=1}^M$. The iterations $m=1,2,\dots, M=1000$ are across the x-axis with the corresponding draw on the y-axis. Results plotted for $t=365$ (top row) and $t=2\cdot 365$ (bottom row). The three chains are illustrated in red, green, black - the default colors in \texttt{coda::plot.mcmc}. Second column: corresponding density plots of the draws $\{ \Psi(t) \}_{m=1}^M$.}
    \label{fig:traceplots}
\end{figure}

When called on objects of class \texttt{mcmc}, such as \texttt{ATE\_chains}, the generic \texttt{plot} function calls the method \texttt{coda::plot.mcmc} to produce Figure \ref{fig:traceplots}. The traceplots on the first column provide evidence of convergence: for each $t$, the three chains converge to the same region regardless of different starting values and do not look autocorrelated. The second column displays the posterior density estimate of $\Psi(t)$ for $t=365$ (top row) and $t=2\cdot 365$ (bottom row) using he draws. Other diagnostics can also be done. For instance, the upper limit of the potential scale reduction factor (PSRF) \citep{Brookes1998} can be computed as:

\begin{lstlisting}
> coda::gelman.diag(ATE_chains)
Potential scale reduction factors:

     Point est. Upper C.I.
[1,]      1.000      1.000
[2,]      0.999      0.999
\end{lstlisting}
An Upper C.I. value near 1 indicates convergence of the chains for each of the two parameters, $\Psi(365)$ and $\Psi(2\cdot 365)$. If the traceplots look autocorrelated and PSRFs are high, then longer warm-ups or more post-warmup draws may improve the quality of the draws.

\begin{figure}[h!]
    \centering
    \includegraphics[width=.5\linewidth]{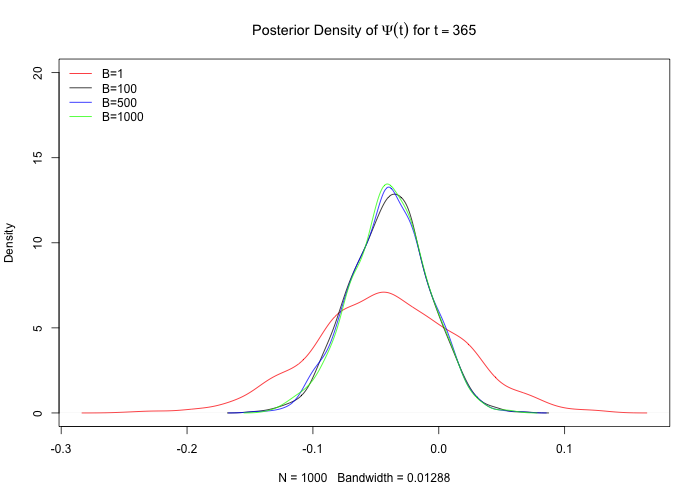}
    \caption{Kernel density visualizations of the posterior for $\Psi(365)$ based on draws $\{\Psi(t)\}_{m=1}^{1000}$ - each obtained under increasing number of Monte Carlo integration iterations, $B$ (see Section \ref{sc:computation}). Clearly $B=1$ is too small an the posterior is much too wide due to the Monte Carlo error in the integration. Increasing to $B=500$ results in a narrower posterior as the Monte Carlo error is eliminated. Increasing to $B=1000$ provides little additional benefit.}
    \label{fig:density}
\end{figure}

Note that in Section \ref{sc:computation}, we additionally require a Monte Carlo integration over the a Bayesian bootstrap draw of the confounder distribution. This was accomplished using $B$ Monte Carlo iterations for each posterior draw. By default, \texttt{bayesgcomp} uses \texttt{B=1000}. For a given set of posterior draws, we can assess if \texttt{B} is sufficiently large by running \texttt{bayesgcomp} with successively larger values of \texttt{B} and checking the resulting posteriors. At some point the posterior will not change much as one increase \texttt{B}, indicating that Monte Carlo error in the integral estimate is minimal. To illustrate, below we obtain 1000 draws from the posterior of $\Psi(365)$ by integrating using $B\in\{1,100,500, 1000\}$ Monte Carlo iterations. Each set of draws under each $B$ is stored in \texttt{ATE\_list}, which we plot in Figure \ref{fig:density}.

\begin{lstlisting}
set.seed(32123)

## post_draws_ar1_adj: results of prev bayeshaz call 

## values of B
B_vec = c(1, 100, 500, 1000)

ATE_list = list() ## shell for storing posterior draws

for(B in B_vec){

        ## run g-comp with B Monte Carlo iterations
        gcomp_res = bayesgcomp(post_draws_ar1_adj,
                               ref = 0, ## treatment reference group
                               B = B, ## monte carlo iterations in g-comp
                               t = 365)
        
        ## store posteriro draws                       
        ATE_list[[length(ATE_list) + 1 ]] = gcomp_res$ATE
}
\end{lstlisting}
As can be seen, the default value of $B=1000$ in this case is sufficient to eliminate the Monte Carlo error in the integration. With $B=1$, the resulting posterior distribution is much too wide as it incorporates variability due both posterior uncertainty as well as Monte Carlo error. In general, the more complicated the covariate distribution the higher $B$ should be used.

\section{Contributing and Issue tracking}

In the paper we describe only the current version of \texttt{causalBETA}, but we imagine that users may want specific extensions and capabilities to be added on later (e.g. other types of priors). Since \texttt{causalBETA} is an open-source \texttt{R} package hosted on GitHub, all users are able (and encouraged) to fork the repository and modify the package as desired. At the time of this manuscript drafting, the package is hosted at \url{https://github.com/RuBBiT-hj/causalBETA}.

Pull requests will be considered to merge these changes to the main package repository as well. Users who notice issues with the software are encouraged to open an issue on the GitHub repository with a reproducible example of their concerns. Bug fixes will be rolled into successive version as time permits.

\newpage

\acks{This work was partially funded the Patient-Centered Outcomes Research Institute (ME-2021C3-24942) and the National Institutes of Health (5R01 AI 167694-3).}


\newpage

\appendix
\section*{Appendix A. Likelihood Construction}
\label{app:theorem}

In this appendix we describe the likelihood for the model described in Section \ref{sc:model} which used in the underlying \texttt{Stan} model to obtain posterior draws of the parameters. As stated in Section \ref{sc:computation}, the likelihood for the observed data is given by
$$ \mathcal{L}( D ; \theta, \beta ) = \prod_{i=1}^n \exp\Big(-\int_{0}^{y_i}  \lambda(u \mid a_i, l_i; \theta, \beta )du \Big )  \lambda(y_i \mid a_i, l_i; \theta, \beta )^{\delta_i} $$

\cite{Laird1981} showed that this likelihood is proportional to the likelihood from a Poisson model fit to the date $D$ after transposing the data to person-time format. Specifically, we denote the last interval at which a patient is in the study be $k_i^*\leq K $. That is, $k_i^*$ is the value such that $I(\tau_{k_i^*-1} < y_i \leq \tau_{k_i^*})=1$. At each interval $k=1, 2,\dots, k_i^*$, we define an event indicator $\delta_{ik} = I( k = k_i^*)$. Clearly $\delta_{ik}=0$ for intervals $k=1,2,\dots, k_i^*-1$ and $\delta_{ik}=1$ at the last interval $k=k_i^*$. We let $\Delta t_{ik}$ be the amount of time a subject was in the study in interval $k$. Recall that each interval has equal length of $\tau_k-\tau_{k-1} = (\max_i y_i) / K := d\tau$. For instance if the maximum time is $\max_i y_i = 365$ days, with $K=100$ intervals each interval will have length $365/100= 3.65$ days. By construction, then, $\Delta t_{ik} = d\tau $ for intervals $k=1,2,\dots, k_i^*-1$ and, in the last interval, the subject is in study for time $\Delta t_{ik_i^*} = y_i - \tau_{k_i^*-1}$. Then, the likelihood above is proportional to

\begin{equation*}
    \mathcal{L}( D ; \theta, \beta ) \propto \prod_{i=1}^n \prod_{k=1}^{k_i^*} \frac{ \exp( - \lambda_{ik} ) ( \lambda_{ik} )^{\delta_{ik}} }{\delta_{ik}! }
\end{equation*}

Where 
$$ \log( \lambda_{ik} ) = \tilde \theta_k + \beta(a, l) + \log( \Delta t_{ik} )$$

This likelihood is easy to specify in standard software. For instance, it can be estimated via MLE using \texttt{glm} function in \texttt{R} by feeding it the transposed data, with outcome $\delta_{ik}$, including a main effect for each interval, and specifying $\log(\Delta t_{ik})$ to be an offset. This is the likelihood used when running \texttt{bayeshaz}.

\vskip 0.2in
\bibliography{sample}

\end{document}